 \RequirePackage{lineno} 
\documentclass[twocolumn,showpacs,preprintnumbers,amsmath,amssymb,prl,superscriptaddress,nofootinbib]{revtex4}

\usepackage{graphicx}
\usepackage{epstopdf}
\usepackage{dcolumn}
\usepackage[loose]{subfigure}      
\usepackage{bm}
\newcommand*{\INFNPi}{\affiliation{INFN Sezione di Pisa$^{a}$; Dipartimento di Fisica$^{b}$ dell'Universit\`a, Largo B.~Pontecorvo~3, 56127 Pisa, Italy}}
\newcommand*{\INFNGe}{\affiliation{INFN Sezione di Genova$^{a}$; Dipartimento di Fisica$^{b}$ dell'Universit\`a, Via Dodecaneso 33, 16146 Genova, Italy}}
\newcommand*{\INFNPv}{\affiliation{INFN Sezione di Pavia$^{a}$; Dipartimento di Fisica$^{b}$ dell'Universit\`a, Via Bassi 6, 27100 Pavia, Italy}}
\newcommand*{\INFNRm}{\affiliation{INFN Sezione di Roma$^{a}$; Dipartimento di Fisica$^{b}$ dell'Universit\`a ``Sapienza'', Piazzale A.~Moro, 00185 Roma, Italy}}
\newcommand*{\INFNLe}{\affiliation{INFN Sezione di Lecce$^{a}$; Dipartimento di Fisica$^{b}$ dell'Universit\`a, Via per Arnesano, 73100 Lecce, Italy}}
\newcommand*{\ICEPP} {\affiliation{ICEPP, The University of Tokyo 7-3-1 Hongo, Bunkyo-ku, Tokyo 113-0033, Japan }}
\newcommand*{\UCI}   {\affiliation{University of California, Irvine, CA 92697, USA}}
\newcommand*{\KEK}   {\affiliation{KEK, High Energy Accelerator Research Organization 1-1 Oho, Tsukuba, Ibaraki 305-0801, Japan}}
\newcommand*{\PSI}   {\affiliation{Paul Scherrer Institut PSI, CH-5232 Villigen, Switzerland}}
\newcommand*{\Waseda}{\affiliation{Research Institute for Science and Engineering, Waseda~University, 3-4-1 Okubo, Shinjuku-ku, Tokyo 169-8555, Japan}}
\newcommand*{\BINP}   {\affiliation{Budker Institute of Nuclear Physics, 630090 Novosibirsk, Russia}}
\newcommand*{\JINR}   {\affiliation{Joint Institute for Nuclear Research, 141980, Dubna, Russia}}
\newcommand*{\ETHZ}   {\affiliation{Swiss Federal Institute of Technology ETH, CH-8093 Z\" urich, Switzerland}}
\newcommand*{\NOVST}  {\affiliation{Novosibirsk State Technical University, 630092, Novosibirsk, Russia}}

\newcommand*{\megsign}        {\mathrm{\mu}^+ \to \mathrm{e}^+ \mathrm{\gamma}}

\newcommand*{\egamma}         {E_{\mathrm{\gamma}}}
\newcommand*{\epositron}      {E_\mathrm{e}}
\newcommand*{\tegamma}        {t_{\mathrm{e \gamma}}}
\newcommand*{\Thetaegamma}    {\Theta_{\mathrm{e \gamma}}}

\newcommand*{\thetaegamma}    {\theta_{\mathrm{e \gamma}}}
\newcommand*{\phiegamma}      {\phi_{\mathrm{e \gamma}}}

\newcommand*{\sens}     { {\cal S}_{90}}
\newcommand*{\ul}     { {\cal B}_{90}}
\newcommand*{\bestfit}     { {\cal B}_\mathrm{fit}}

\begin{document}
\title{New constraint on the existence of the   $\megsign$  decay  }
\author{J.~Adam}               \PSI\ETHZ
\author{X.~Bai}                \ICEPP
\author{A.~M.~Baldini$^{a}$}   \INFNPi
\author{E.~Baracchini}         \ICEPP\KEK\UCI
\author{C.~Bemporad$^{ab}$}    \INFNPi
\author{G.~Boca$^{ab}$}        \INFNPv
\author{P.~W.~Cattaneo$^{a}$}  \INFNPv
\author{G.~Cavoto$^{a}$}       \INFNRm
\author{F.~Cei$^{ab}$}         \INFNPi
\author{C.~Cerri$^{a}$}        \INFNPi
\author{A.~de~Bari$^{ab}$}      \INFNPv
\author{M.~De~Gerone$^{ab}$}   \INFNGe
\author{T.~Doke}               \Waseda
\author{S.~Dussoni$^{a}$}     \INFNPi
\author{J.~Egger}              \PSI
\author{Y.~Fujii}              \ICEPP
\author{L.~Galli$^{a}$}       \PSI \INFNPi
\author{F.~Gatti$^{ab}$}        \INFNGe
\author{B.~Golden}              \UCI
\author{M.~Grassi$^{a}$}        \INFNPi
\author{A.~Graziosi$^{a}$}           \INFNRm
\author{D.~N.~Grigoriev}        \BINP \NOVST
\author{T.~Haruyama}            \KEK
\author{M.~Hildebrandt}         \PSI
\author{Y.~Hisamatsu}           \ICEPP
\author{F.~Ignatov}             \BINP
\author{T.~Iwamoto}             \ICEPP
\author{D.~Kaneko}             \ICEPP
\author{P.-R.~Kettle}           \PSI
\author{B.~I.~Khazin}           \BINP
\author{N.~Khomotov}             \BINP
\author{O.~Kiselev}             \PSI
\author{A.~Korenchenko}         \JINR
\author{N.~Kravchuk}            \JINR
\author{G.~Lim}              \UCI
\author{A.~Maki}                \KEK
\author{S.~Mihara}              \KEK
\author{W.~Molzon}              \UCI
\author{T.~Mori}                \ICEPP
\author{D.~Mzavia\stepcounter{footnote}}              \JINR
\author{R.~Nard\`o}             \INFNPv
\author{H.~Natori}              \KEK\ICEPP\PSI
\author{D.~Nicol\`o$^{ab}$}     \INFNPi
\author{H.~Nishiguchi}          \KEK
\author{Y.~Nishimura}           \ICEPP
\author{W.~Ootani}              \ICEPP
\author{M.~Panareo$^{ab}$}      \INFNLe
\author{A.~Papa}                \PSI
\author{G.~Piredda$^{a}$}       \INFNRm
\author{A.~Popov}               \BINP
\author{F.~Renga$^{a}$}         \INFNRm\PSI
\author{E.~Ripiccini$^{ab}$}         \INFNRm
\author{S.~Ritt}                \PSI
\author{M.~Rossella$^{a}$}      \INFNPv
\author{R.~Sawada}              \ICEPP
\author{F.~Sergiampietri$^{a}$}\INFNPi
\author{G.~Signorelli$^{a}$} \INFNPi
\author{S.~Suzuki}              \Waseda
\author{F.~Tenchini$^{ab}$}     \INFNPi
\author{C.~Topchyan}            \UCI
\author{Y.~Uchiyama}            \ICEPP\PSI
\author{C.~Voena$^{a}$}         \INFNRm
\author{F.~Xiao}                \UCI
\author{S.~Yamada}              \KEK
\author{A.~Yamamoto}            \KEK
\author{S.~Yamashita}           \ICEPP
\author{Z.~You}           \UCI
\author{Yu.~V.~Yudin}           \BINP
\author{D.~Zanello$^{a}$}       \INFNRm

\collaboration{MEG Collaboration}
\noaffiliation

\date{\today}

\begin{abstract}
The analysis of a combined dataset, totaling $3.6\times 10^{14}$ stopped muons on target, in the search for the lepton flavour violating decay  $\megsign$ is presented. The data collected by the MEG experiment at the Paul Scherrer Institut show no excess of events compared to background expectations and yield a new upper limit on the branching  ratio of this decay of $5.7\times10^{-13}$ (90\% confidence level). This represents  a four times more stringent limit than the previous world best limit set by MEG.

\end{abstract}


\maketitle
The lepton flavour violating $\megsign$ decay
is predicted to have an unobservable low rate within the Standard Model of elementary particle 
physics (SM), despite the existence of neutrino oscillations\,\cite{petcov}. 
Conversely, the majority of new physics 
models\,\cite{barbieri, hisano, LFV-EPC, isidori} Beyond SM (BSM),
particularly in view of the  recent 
measurements of a large  $\theta_{13}$ at reactor\,\cite{DayaBay, Reno, DoubleChooz} 
and accelerator\,\cite{T2K} experiments, predict measurable rates  for this decay. 
An observation of  the 
$\megsign$ decay would therefore represent an unambiguous sign of BSM physics, 
whereas  improvements in the branching ratio upper limit constitute significant 
constraints on the parameter space, complementary to those 
obtainable at high energy colliders. 

The present best upper limit on the $\megsign$ decay branching ratio ${\cal{B}}$
(${\cal{B}}  < 2.4 \times 10^{-12}$ at 90\% C.L.)
was set by the MEG experiment \,\cite{MEG2011} with an analysis of  the data taken  in
the years 2009--2010, for a total number of $1.75\times 10^{14}$ positive muons stopped 
on target.

In this paper we present an updated analysis of the 2009--2010 data 
sample, based on recently  improved algorithms for the reconstruction 
of positrons and photons together with the analysis of the data sample collected 
in 2011 with a beam intensity of $3 \times 10^{7} \mu^{+}/{\rm s}$, 
which corresponds to $1.85 \times 10^{14}$ stopped muons on 
target. 
Furthermore the combined analysis of the full 2009--2011 statistics is  presented. 

The signature of the signal event is given by a back-to-back, monoenergetic, time coincident 
photon-positron pair from the two body $\megsign$ decay.
In each event, positron and photon candidates are described by five 
observables: the photon and positron energies ($\egamma$, $\epositron$), 
their relative directions ($\thetaegamma$, $\phiegamma$)\,\cite{angledef}
and emission time ($\tegamma$). 
Our analysis is based on a maximum likelihood technique
applied in the analysis region defined by
$48<\egamma<58\,$MeV, $50<\epositron<56\,$MeV,
$\left|\tegamma\right|<0.7\,$ns,
$\left|\thetaegamma\right|<50\,$mrad and
$\left|\phiegamma\right|<50\,$mrad, which is described in 
detail in \cite{MEG2011}.
We call \lq\lq time sidebands\rq\rq ~the regions in the variable 
space defined by 1\,$<\left|\tegamma\right|<4\,$ns, 
\lq\lq $\egamma$-sideband\rq\rq~that defined by $40<\egamma<48~{\rm MeV}$ 
and \lq\lq angle sidebands\rq\rq~those defined by 
50\,$<\left|\phiegamma\right|<150\,$mrad or 
50\,$<\left|\thetaegamma\right|<150\,$mrad. 
 
The background has two components, one coming 
from the Radiative Muon Decay $\mathrm{\mu}^{+}\rightarrow{\rm e}^{+} 
\mathrm{\nu} \bar{\mathrm{\nu}} \mathrm{\gamma}$ (RMD) and one from the accidental superposition 
of energetic positrons from the standard muon Michel decay with photons 
from RMD, positron-electron annihilation-in-flight or bremsstrahlung. 
At the MEG data taking rate, 93\% of events with $E_{\mathrm{\gamma}}>48~{\rm MeV}$ are from the ACCidental background (ACC).

The MEG detector is described in detail elsewhere\,\cite{techpaper}. It is comprised of
a positron spectrometer formed by a set of Drift CHambers (DCH) and scintillation Timing Counters 
(TC), located inside a superconducting solenoid with a gradient magnetic field along the
beam axis, and a photon detector,
located outside of the solenoid, made up  of a homogeneous volume ($900\,\ell$)
of Liquid Xenon (LXe) viewed by 846 UV-sensitive photomultiplier tubes (PMTs) submerged in the liquid.

The MEG detector response, resolutions and stability are constantly monitored and calibrated
by means of a multi-element calibration system\,\cite{techpaper, MEG2011},
which was recently complemented with a new method based on the Mott scattering of a monochromatic positron beam. 
Two important hardware improvements were introduced in 2011, 
one in  the $\mathrm{\pi}^{-} {\rm p} \rightarrow \mathrm{\pi}^{0} {\rm n}$ Charge EXchange reaction (CEX) calibration by 
replacing the NaI detector, used to define the back-to-back coincidence with 
the LXe detector when observing the two photons from $\mathrm{\pi}^{0}$ 
decay, with a higher resolution BGO array detector, the other in the optical survey technique for the DCH 
by using a laser tracker and prismatic corner cube reflectors mounted on the DCH modules.

In 2011 the DAQ efficiency, being the product of DAQ live time fraction 
and trigger efficiency to select signal event, was at the level of 96\%. An 
improvement from 72\% in 2010 is due to a new multiple buffer read out 
scheme, which guarantees a 99\% live time fraction with a less 
constrained event selection, thus allowing a 97\% signal trigger
efficiency\,\cite{trigger}.

The positron track is reconstructed by combining its measured positions at 
each DCH layer (hit) in the spectrometer. 
Longitudinal (along the muon beam direction) $z$-positions are derived from  signals induced on the 
segmented DCH cathodes. 
Electromagnetic noise has been the main source of degradation of the $z$-resolution during all the data-taking periods and a new 
reconstruction algorithm based on a fast Fourier transform filtering 
technique has been applied to mitigate such effect, yielding an up to $\sim$10\% improvement
in angular resolution. Internal alignment of the DCH layers is obtained by tracking  cosmic ray muons   with the magnet off and using an optical survey, as described in \cite{MEG2011}.
              

   The positron kinematic variables are extracted by means of a Kalman filter track fitting
technique\,\cite{Kalman1,Kalman2}.
    This algorithm has been completely revised for this analysis to include a better 
model for the hits and the track itself, based on the GEANE package\,\cite{GEANE1,GEANE2}.
An improved  model for the detector material, accounting for 
multiple scattering and energy loss as well as a detailed map of the non-uniform magnetic field, 
measured with a 0.2\% precision has also been included.
The fitted positron track is propagated to the TC allowing an 
iterative refinement of the hits with the positron time measurement.

The track fit yields a covariance matrix for the parameters, resulting in very good agreement 
with the measured resolutions, which are extracted from a sample of tracks 
with two full turns in the DCH, by comparing the track parameters determined independently for each turn at an intermediate plane.
Consequently a per-track error is determined which allows us to follow the variable 
DCH performance during the data-taking period and is taken 
into account in the maximum likelihood analysis.

\begin{figure}
\centering
\includegraphics[clip, width=.9\columnwidth]{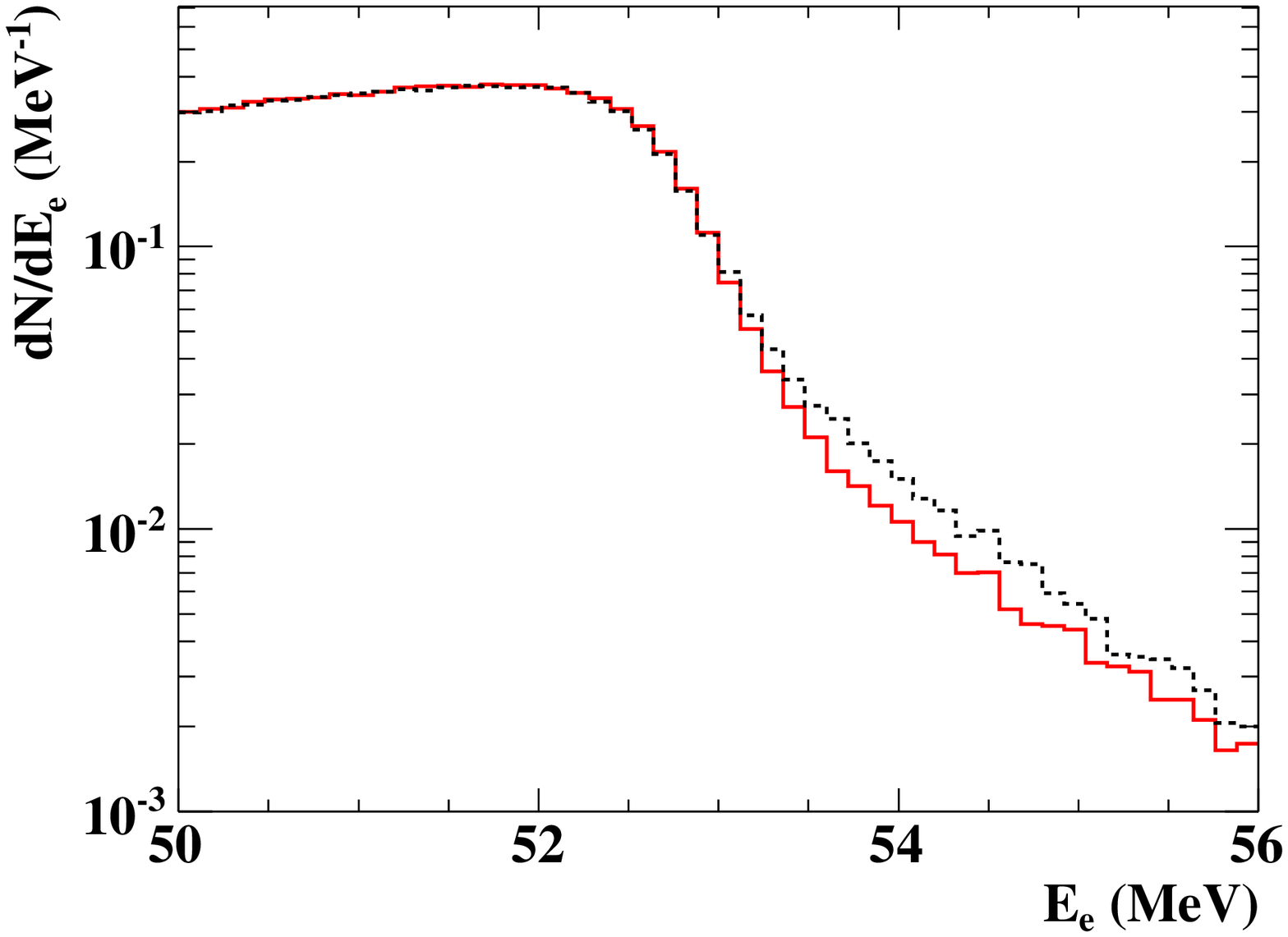}
\includegraphics[clip, width=.9\columnwidth]{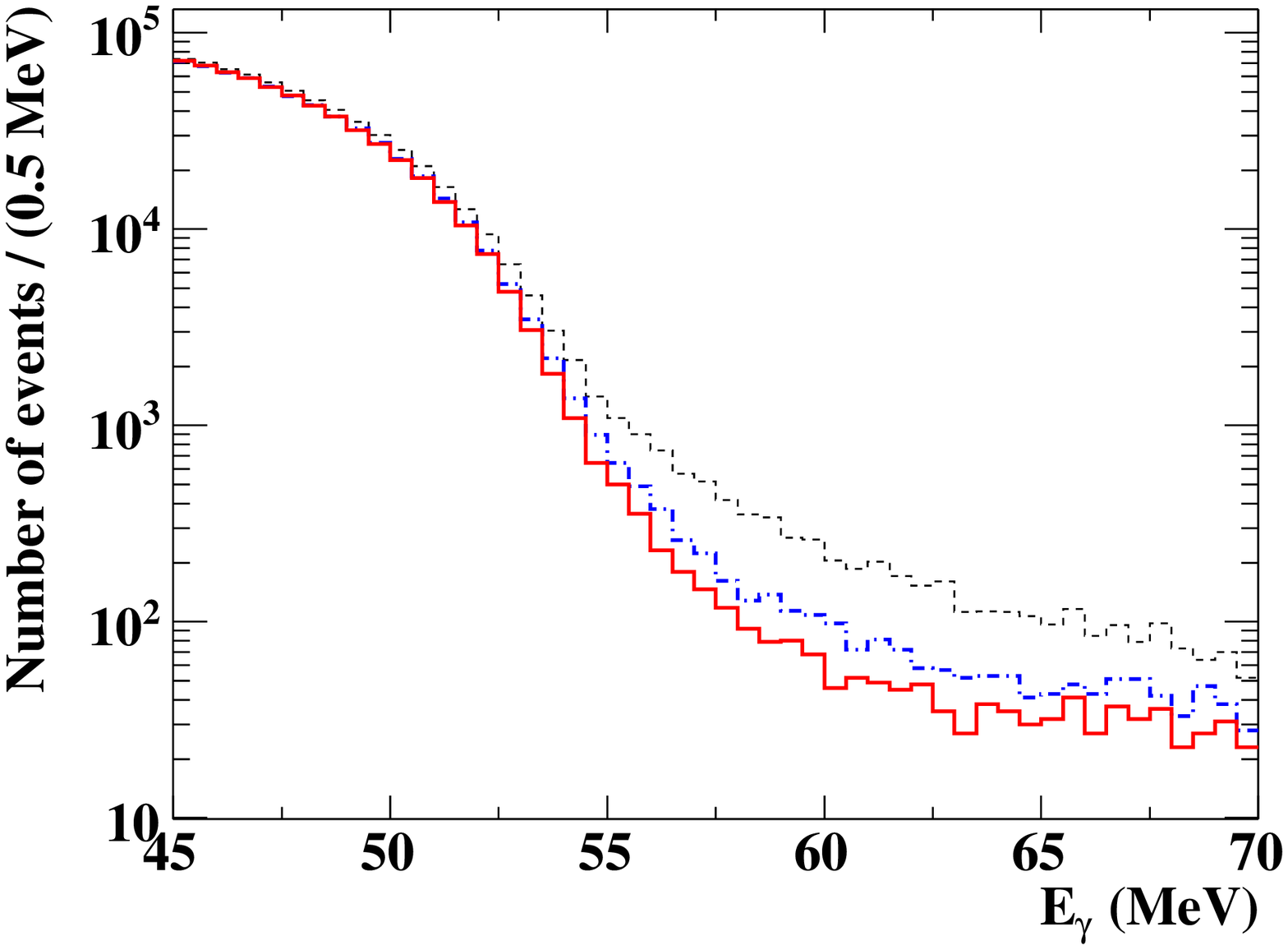}
\begin{center}
\caption{
(Top) The Michel positron  spectrum in the same dataset in the time sidebands with the old 
(black dashed line)  and the new (red solid line) track reconstruction code. 
(Bottom) The photon background spectra from the time sidebands of the muon 
data in 2010 with different pile-up elimination algorithms. Black dots: no pile-up elimination;  blue dot-dashed:  previous algorithm;  red solid: new algorithm.}
\label{fig:improve}
\end{center}
\end{figure}

The average hit multiplicity for a track is about 10 and only tracks with at least 
7 hits and either one or two turns in the 
spectrometer are retained for the analysis. Additional quality requirements 
on the  $\chi^2$ fit value and parameter uncertainties are applied to select only one 
positron per event. The overall improvement in positron 
reconstruction with respect to the previous data analysis is clearly visible
in Fig.\ref{fig:improve} (top) where the reconstructed positron 
energy near the kinematic edge of the Michel decay spectrum 
shows a reduced tail. The energy resolution is
evaluated  by fitting the kinematic edge and it is well described by 
the sum of three Gaussian curves with a resolution of $\sigma_{E_\mathrm{e}}$ = 305 \,keV 
for the core component (85\%).         

The resolution in the azimuthal angle of the positron when it leaves the target, $\phi_\mathrm{e}$,
has a dependence on $\phi_\mathrm{e}$ itself with a minimum at 
$\phi_\mathrm{e}$ = 0, where it is measured by the two-turn 
method after correcting for all known correlations to be $\sigma_{\phi_{\mathrm e}}$ = 7.5 (7.0)\,mrad \cite{bracketnumbers}. 
Similarly, the resolution in the polar angle, $\theta_{\mathrm e}$, 
is measured to be $\sigma_{\theta_{\mathrm e}}$ = 10.6 (10.0)\,mrad.
The decay vertex coordinates and the positron direction at the vertex are 
determined by extrapolating the reconstructed track back to the target. 
The resolutions on the decay vertex coordinates are also determined by the two-turn 
method  and are described by a Gaussian curve with 
$\sigma_{z}$ = 1.9 (1.5)\,mm and, in the vertical direction,
by the sum of two Gaussian curves with $\sigma_{y}$ = 1.3 (1.2)\,mm 
for the core component (85\%).

The LXe detector uses the xenon scintillation light to measure the total energy released by the photon, 
the position and time of its first interaction. 
The three-dimensional photon conversion point is reconstructed by using 
the distribution of the number of scintillation photons detected by the PMTs near the incident position.
The photon direction is defined by the line connecting the decay 
vertex to the photon conversion point in the LXe detector.
The reconstruction of the photon energy is based on the sum of the scintillation light detected by all PMTs. 
Monochromatic 55\,MeV photons from $\mathrm{\pi}^{0}$ decays are used to determine the absolute energy scale.
The photon conversion time is reconstructed by combining the leading edge times 
of the PMT waveforms. 

It is important to identify and unfold photon pile-up events at high muon rates 
since at $3 \times 10^{7} 
\mu^{+}/{\rm s}$ beam rate, around 15\% of triggered events suffer from 
pile-up. For the previously published analyses,  photon pile-up 
events were identified topologically by the pattern of PMT light distribution 
and temporally by the leading edge time distribution in the time reconstruction, 
without the use of detailed waveform information.
In addition to these methods,  a new algorithm, analyzing waveforms 
after summing up all channels at the end of the full chain of photon reconstruction,
was developed.
It enables the efficient  identification and removal of
pile-up photons by using template waveforms.
Consequently the charge integration window for the energy estimate is re-adjusted, 
resulting in a better energy reconstruction.
This improvement is shown in Fig.~\ref{fig:improve} (bottom) 
where we compare the photon energy spectra obtained
with different pile-up elimination algorithms. 
The reduced tail in the high energy region is clearly 
visible. The efficiency of photon reconstruction is improved from 
59\% to 63\% due to the new algorithm.

The performance of the position reconstruction is evaluated by a Monte Carlo
simulation, resulting in resolutions of  5\,mm on the photon entrance face, 
and 6\,mm along the radial depth from the entrance face.
This is validated in CEX runs by placing lead slit collimators in front of the LXe detector. 
The timing and the energy resolutions are evaluated using two simultaneous back-to-back 
photons from $\mathrm{\pi}^{0}$ decay. 
The LXe timing resolution is found to be $\sigma_{t_{\mathrm{\gamma}}}$ = 67\,ps at the signal energy.
The position-dependent energy resolutions are measured in the CEX data
and the average energy resolution extracted from a Gaussian fit to the high energy side 
of the spectrum is evaluated to be 1.7\% (1.9\%) and 2.4\% 
(2.4\%) for radial depths larger and smaller than 2 cm 
respectively. 
These position-dependent energy resolutions are incorporated into the likelihood analysis.

The resolutions of the relative directions ($\thetaegamma$, $\phiegamma$) are derived 
by combining the relevant resolutions of positrons and photons discussed above.
The results are
16.2 (15.7)\,mrad for $\thetaegamma$ and 8.9 (9.0)\,mrad for $\phiegamma$ after correcting for all known correlations.
The relative time $\tegamma$ is derived from the time measurements, one in 
the LXe detector and the other in the TC, after correcting for the lengths of the particle 
flight-paths. The associated resolutions at the signal energy 
are 127 (135)\,ps, evaluated from the RMD peak observed in the 
$\egamma$-sideband; a small correction takes into account the $\egamma$-dependence
measured in the CEX calibration runs.
The position of the RMD-peak corresponding to $\tegamma$ = 0 was 
monitored constantly during the physics data-taking period and found 
to be stable to within $15$\,ps. 

A blind analysis procedure is applied only to the new dataset in 2011 by masking a region of 
$48<\egamma<58$\,MeV and $\left| \tegamma
\right|< 1\,{\rm ns}$
until the Probability Density Functions (PDFs) for the likelihood function are finalized. 
For all the datasets including 2009--2010 data the background studies and the extraction of the PDFs are carried out in the time and angle sidebands.
The maximum likelihood fit is performed in order 
to estimate the number of signal, RMD and ACC events in the analysis region.
The definition of the likelihood function is described in detail in \cite{MEG2011}.
All PDFs as a function of the observables are extracted from the data.
Different resolutions and correlations are used in the PDFs
on an event-by-event basis. 
The dependence on the photon interaction position
and the quality of the positron tracking has already been incorporated into the 
previous analysis, while in the new analysis a per-event error matrix for the 
positron observables, estimated by the new Kalman filter, has been introduced into 
the PDFs. 
The sensitivity is improved by about 10\% in the new analysis with the positron per-event error matrix.
An analysis with constant PDFs is also performed as a crosscheck, 
showing consistent results.
The confidence interval for the number of signal events is calculated by a 
frequentist method with a profile likelihood-ratio 
ordering\,\cite{MEG2011, PDG, Feldman-Cousins}, 
where the numbers of RMD and ACC events are treated as nuisance parameters.

To translate the estimated number of signal events into a signal branching ratio two independent normalization methods 
are used, either counting the number of Michel positrons selected with a dedicated trigger or the number of RMD events observed in the muon data.
Their combination leads to a 4\% uncertainty in the branching ratio estimate.
The increased reconstruction efficiency of the new algorithms results in a 14\% larger 
data sample for the $\megsign$ search, as estimated with both normalization methods, with both positron and photon new algorithms contributing equally.

The systematic uncertainties on the PDF parameters and 
on the normalization are taken into account 
in the calculation of the confidence intervals
by fluctuating the PDFs by the 
amount of the uncertainties. In total they produce 
a 1\% effect on the observed upper limit, with the majority of the 
contribution coming from the angular PDFs.

The sensitivity ($\sens$) is estimated as the median of the distribution 
of the branching ratio upper limits at $90\%$ C.L.,
calculated over an ensemble of pseudo-experiments, randomly 
generated according to the PDFs based on a 
null signal hypothesis, with the rates of ACC and RMD 
evaluated from the sidebands. The sensitivities have been so evaluated  
for the 2009--2010 combined data, the $2011$ 
data alone and the 2009--2011 combined data sample, and are reported in Table \ref{tab:BRtable}. 
Likelihood analyses are also performed in fictitious analysis regions
in both the time- and angle-sidebands, getting upper limits  all 
in good agreement with the $\sens$'s.

\begin{figure}[htb]
\centering
\includegraphics[clip, width=.9\columnwidth]{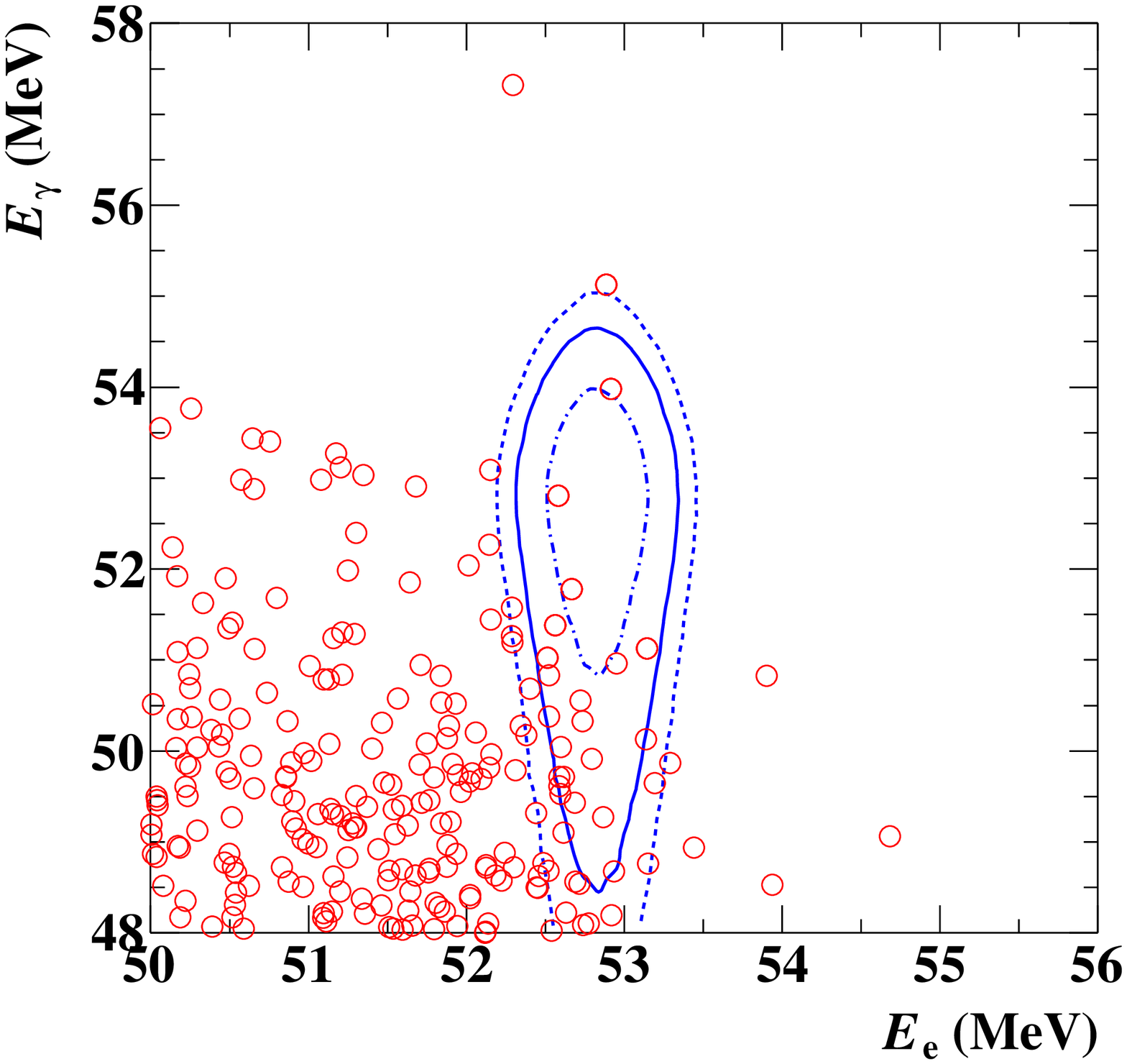}
\includegraphics[clip, width=.9\columnwidth]{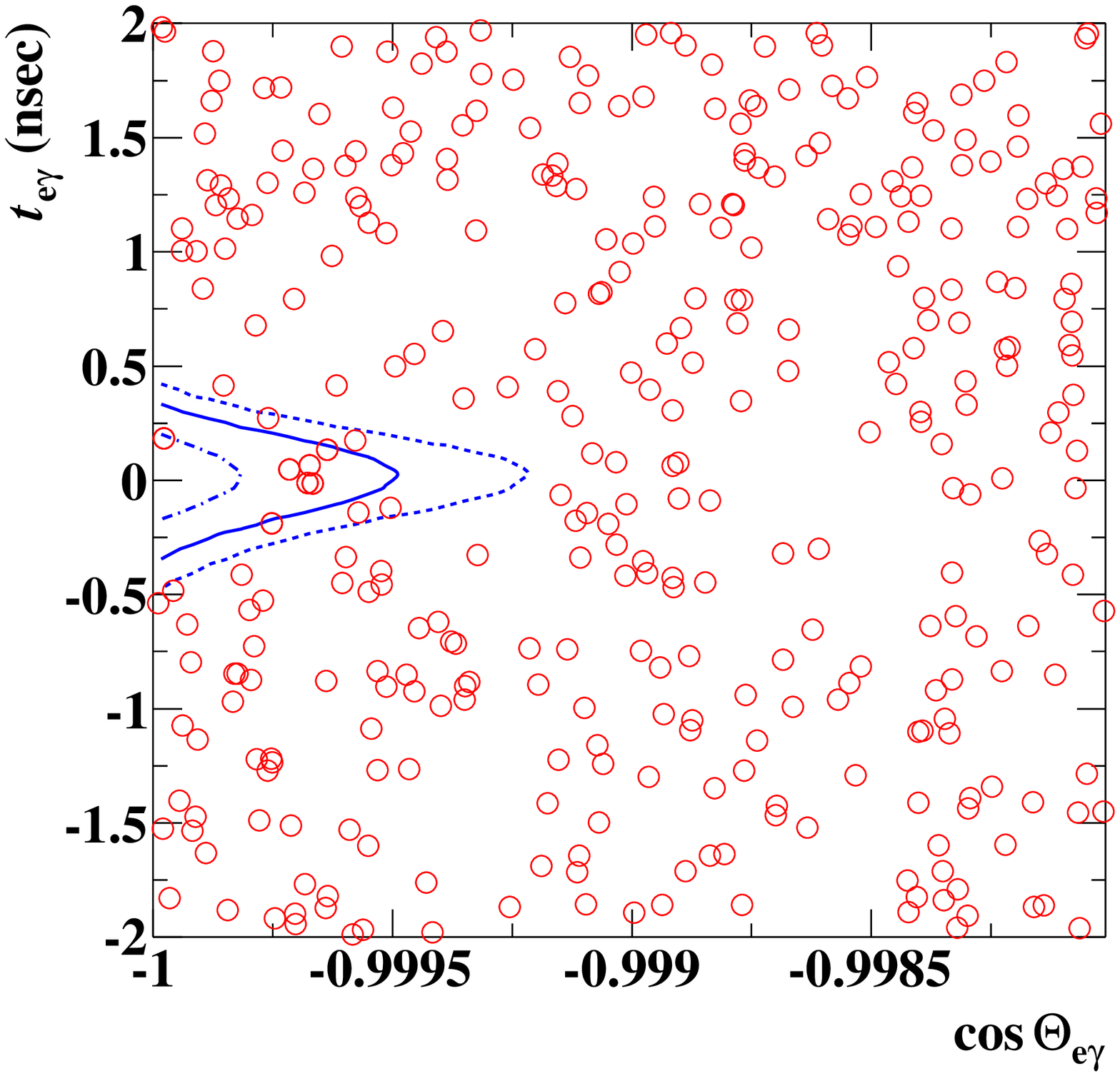}%
\caption{\label{fig:event distribution}
Event distributions for the combined 2009--2011 dataset 
in the ($\epositron, \egamma$)- and ($\cos\Thetaegamma, \tegamma$)-planes. 
In the top (bottom) panel, a selection of 
$|\tegamma|<0.244\,{\rm ns}$ and $\cos\Thetaegamma<-0.9996$
with 90\% efficiency for each variable
($52.4<\epositron<55\,{\rm MeV }$ and $51<\egamma<55.5\,{\rm MeV}$ 
with 90\% and 74\% efficiencies for $\epositron$ and $\egamma$, respectively)
is applied.
The signal PDF contours (1, 1.64 and 2 $\sigma$) are
also shown.
}
\end{figure}

Figure \ref{fig:event distribution} shows the event distributions 
in the ($\epositron, \egamma$)- and ($\cos\Thetaegamma, \tegamma$)-planes 
for the combined 2009--2011 dataset, where $\Thetaegamma$ is 
the opening angle between positron and photon, 
together with the contours of the averaged signal PDFs. 

\begin{figure}[tb]
\begin{center}
\includegraphics[width=.9\columnwidth, clip=true]{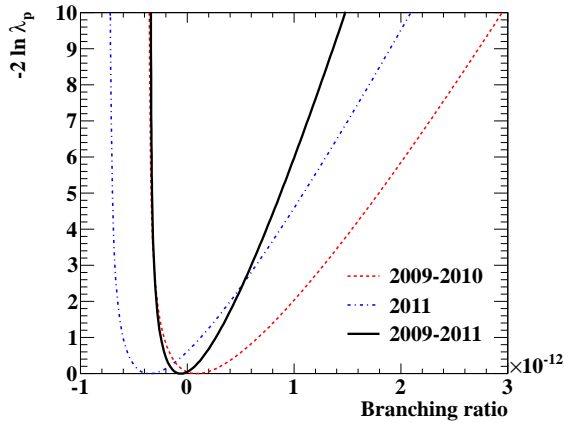}
\caption{\label{fig:profile likelihood ratio}
Observed profile likelihood ratios ($\lambda_\mathrm{p}$) as a function of the branching ratio
for the 2009--2010 combined data, the $2011$ data alone 
and the combined 2009--2011 data sample.}
\end{center}
\end{figure}

The observed profile likelihood ratios as a function of the 
branching ratio are shown in Fig.\,\ref{fig:profile likelihood ratio}.
The best ${\cal{B}}$ estimates, upper limits at 90\% C.L.
($\ul$) and $\sens$ 
for the combined 2009--2010 dataset, the 2011 data alone 
and the total 2009--2011 dataset are listed
in Table\,\ref{tab:BRtable}. 
The $\ul$
for the latter is $5.7 \times 10^{-13}$. 
As a quality check the maximum likelihood fit is repeated for the 2009--2011 dataset 
omitting the constraint on the number of background events.
We obtain $N_{\mathrm{RMD}}$ = 163 $\pm$ 32 and $N_{\mathrm{ACC}}$ = 2411 $\pm$ 57,
in good agreement with the expectations estimated from $\egamma$ and time sidebands, 
$\left<N_{\mathrm{RMD}}\right>$ = 169 $\pm$ 17 and 
$\left<N_{\mathrm{ACC}}\right>$ = 2415 $\pm$ 25.

\begin{table}[htb]
   \caption{\label{tab:BRtable} 
   Best fit values (${\cal B}_\mathrm{fit}$'s), branching ratios ($\ul$) and sensitivities ($\sens$)}
\begin{center}
\begin{tabular}{lccc}
\\{\bf Dataset} & $\bestfit\times10^{12}$\,\, & $\ul\times10^{12}$\,\, & $\sens\times10^{12}$ \\[1mm] 
\hline
\hline\\[-3mm]

2009--2010           & $\hspace*{\fill} 0.09$ \;\;\;\;\;     & $1.3$ \,     & $1.3$ \\
2011                & $\hspace*{\fill} -0.35$\;\;\;\;\;\,\, & $0.67$\,\,\, & $1.1$ \\
2009--2011 \,\, & $\hspace*{\fill} -0.06$\;\;\;\;\;\,\, & $0.57$\,\,\, & $0.77$ \\

\hline \hline
\end{tabular}
\end{center}
\end{table}

The reanalysis of the 2009--2010 dataset with new algorithms has led to variations in 
the values of the observables which are much smaller than the detector resolutions.
The events observed with the highest signal-likelihood in the previous analysis of the 2009--2010 dataset
have also moved in the new analysis within the expected fluctuation and mostly 
have had their signal-likelihood slightly reduced.
These small variations induce a change 
in $\bestfit$ and $\ul$ for the same dataset.
We have compared $\ul$'s obtained with the new and old analyses for the same sample 
of simulated experiments and found that a change of $\ul$ equal to or larger
than what we observe in the 2009--2010 dataset has a 31\% probability of occurring.
The upper limit obtained from the 2011 data only is more stringent than 
$\sens$. This is, however, not considered unusual, 
since the probability to have $\ul$ equal or smaller than that observed in the 2011 
data is calculated to be 24\% with a sample of simulated experiments.

In conclusion the MEG experiment has so established the most stringent upper 
limit to date on the branching ratio of the $\megsign$ decay, 
${\cal{B}} < 5.7\times 10^{-13}$ at $90\%$ C.L. using data collected between 
$2009$ and $2011$,  
which improves the previous best upper limit by a factor of four.
Further  data  have also  been acquired in $2012$ with  an additional three-month run scheduled for $2013$; the final number of 
stopped muons is expected to be almost twice that of  the sample analyzed so far.
Currently an upgrade program is underway  aiming at a sensitivity improvement of a further 
order of magnitude \cite{Baldini:2013ke}.  

\section{Acknowledgements}                                                                                   
We are grateful for the support and cooperation provided by PSI 
as the host laboratory and to the technical and engineering staff of our institutes.
This work is supported by SNF grant 137738 (CH), DOE DEFG02-91ER40679 (USA), INFN (Italy) and 
MEXT KAKENHI 22000004 (Japan). Partial support of the Italian Ministry 
of University and Research (MIUR) grant RBFR08XWGN is acknowledged.

\end{document}